# Experimental Observation of Mesoscopic Dynamic Fluctuations in Water-Alcohol Mixtures

Yukio Kajihara[1)], Nanako Shibata[2)] and Masanori Inui[1)]

## AFFILIATIONS

[1)]Graduate School of Advanced Science and Engineering, Hiroshima University, Higashi-Hiroshima, Hiroshima, 739-8521, Japan

[2)]Graduate School of Integrated Arts and Sciences, Hiroshima University, Higashi-Hiroshima, Hiroshima, 739-8521, Japan

## ABSTRACT

We applied a dynamic fluctuation measurement method to liquid water–methanol mixtures over a wide range of compositions and temperatures. This method, which we developed in recent years, combines ultrasonic and inelastic X-ray scattering techniques to detect mesoscopic-level fluctuations—that is, the degree of heterogeneity—in liquids using sound waves. As a result, it was found that dynamic fluctuations exist across a wide range of compositions and temperatures, including pure substances (water and methanol). Furthermore, in the low-methanol composition range (approximately 20% mol or less), the strength of dynamic fluctuation increased toward lower compositions and lower temperatures, suggesting that the origin of the fluctuation lies in the low-temperature (supercooled) region of water alone. Conversely, this indicates that in the range below 20% methanol, fluctuations in water alone are dominant rather than the mixed effects of water and methanol. We performed similar dynamic fluctuation measurements on water–ethanol and water–glycerol mixtures at room temperature. We found that dynamic fluctuations exist universally across the entire composition range, and confirmed an increase in the strength of dynamic fluctuations toward lower compositions in the regions approximately below 15% and below 20%, respectively. This revealed that the phenomenon is not unique to methanol systems but is a common behavior shared by water–alcohol mixtures. It has long been known that water-alcohol mixtures exhibit various thermodynamic anomalies, but the mechanisms behind them have not yet been fully elucidated. The present results are expected to provide a new direction for discussion on this issue.

## I. INTRODUCTION

We have recently developed a new "dynamic fluctuation" measurement method [1, see also section II.A]; this method combines two sound velocity measurement techniques with vastly different frequencies—inelastic X-ray scattering (IXS, THz frequency range) and ultrasonic measurement (US, MHz)—to quantify the magnitude of the frequency dependence of the sound velocity as the strength of the fluctuations. We have applied this technique to liquid water and have demonstrated for the first time that "dynamic fluctuations" are universally present in liquids across nearly the entire pressure-temperature range, and that their strength is particularly high at low temperature and low pressure. It is generally recognized that liquids in thermal equilibrium above their melting points are homogeneous, except “in fluids near a critical or convolution point” [2]. Therefore, the fact that these “dynamical fluctuation” exists universally represents a new insight that must be considered when understanding the properties of liquids. In fact, the temperature- and pressure-dependence of the dynamic fluctuation strength in liquid water correlates over a wide range with the temperature- and pressure-dependent changes in isochoric specific heat, indicating that strength of dynamic fluctuation is an important physical quantity in discussions of liquid thermodynamics. To further explore the relationship between dynamic fluctuations and thermodynamics in liquids, it is desirable to apply this method to other systems.

It has long been known that water–alcohol mixtures exhibit various thermodynamic anomalies. For example, it is known that there are minima and maxima in the composition dependence of excess enthalpy and excess molar volume [3,4], as well as a maximum in the composition dependence of the ultrasonic sound velocity [5,6,7], and that the composition at which this maximum occurs varies depending on the type of alcohol. Interpretations of these thermodynamic anomalies in water–alcohol mixtures vary widely [3-7]; the presence of a rigid clathrate structure seems plausible as an explanation for the maximum in the composition dependence of the sound velocity [5,6]; alternatively, hydrogen bond is a concept frequently used when discussing the structure and physical properties of water and alcohols, and many discussions focus on the dimensionality of the hydrogen-bonding networks in water and alcohol, as well as concepts such as hydrophobic hydration [8,9]. In anyway, no comprehensive interpretation has been reached.

In this study, we apply the above dynamic fluctuation measurement method to this water-alcohol mixture systems and discuss the thermodynamic anomalies of the system from the perspective of fluctuations. Such an approach has not been taken before.

## II. EXPERIMENTAL

### A. Principle of Dynamic Fluctuation Measurement Method

First, we explain the principle of the dynamic fluctuation measurement method using sound waves. Generally, when some form of "fluctuation" (temporal or spatial inhomogeneity or heterogeneity in mesoscopic scales) exists within a sample as shown in Fig. 1 (a), sound waves are scattered by that fluctuation, resulting in a decrease in their effective sound velocity. In the past, this principle was used to actually measure critical fluctuations associated with liquid–gas phase transitions (LGT) in various liquids [10,11], which made a significant contribution to the theoretical development of critical phenomena [12]. However, the sound velocity actually decreases only when the frequency ($\omega$) of the sound wave is comparable to or lower (slower) than the characteristic frequency of the fluctuations (the reciprocal of characteristic time) of the sample; when the frequency is sufficiently high, the sound velocity does not decrease, and the "bare sound velocity"—unaffected by the fluctuations and thus directly corresponds to the stiffness of the local structure—is measured. This is illustrated in Fig. 1 (b). In other words, by measuring the sound velocity in sufficiently low-frequency bands (typically US, MHz range) and sufficiently high-frequency bands (IXS, THz), it is possible to measure the "strength of dynamic fluctuation" as,

$$\Delta_v \equiv \frac{v\left(\omega=\infty\right) - v(\omega = 0)}{v(\omega = 0)} \approx \frac{v_{IXS} - v_{US}}{v_{US}} \tag{1}$$

Although the frequency dependence of fluctuations and the sound velocity had already been recognized in the 1950s—during the early stages of ultrasonic research on critical phenomena in LGT [10]—it was not until the 1990s, when IXS measurements using third-generation synchrotron facilities became possible, that it became feasible to measure the sound velocity at sufficiently high frequencies. We were the first to experimentally extract this dynamic fluctuation strength [1,13,14], and it can be said to represent a new physical quantity and concept that has not been discussed previously.

### B. Ultrasonic Measurement

The measurements were performed with a standard pulse echo technique. The ultrasonic measurement system was constituted by Tsuchiya and details of the equipment are reported in [15]. A piezoelectric ceramic transducer operating at about 10MHz was used to produce and detect ultrasonic pulse waves. A quartz cell was used as a container of a liquid sample. The cell has a quartz buffer rod for sound waves to travel between the sample and transducer. A length of the sample thickness was about 6mm, and the exact sample length was determined by measuring the sound velocity of distilled water and calibrated by reported values at 298 K in [16]. Note that the thermal expansion coefficient of quartz is $\alpha = 0.5 \times 10^{-6}$ [/K]. Within the temperature range of this experiment (about 150 K), the sample length could change by approximately 0.5 microns; this is a negligible, and no correction was applied. A sound propagation time between a delay line end and a reflector was measured using a time function of an oscilloscope (LeCroy LT262). For the samples used in this study, this propagation time was approximately 6–12 $\mu$s, the measurement accuracy was better than 0.01 $\mu$s; consequently, the measurement accuracy of the sound velocity is approximately 1–3 [m/s]. To achieve temperature condition at the sample position from 178K to 323 K, a refrigerator (Twinbird SC-UE15R) and handmade Nichrome heater were introduced, while the transducer was kept at an ambient temperature. The temperatures and compositions of the sample for US measurements are shown by blue circles in Fig. 2.

### C. Inelastic X-ray Scattering Measurement

The IXS experiments were carried out at the high-resolution inelastic X-ray scattering beam line BL35XU at SPring-8 in Japan. Backscattering at the Si (11 11 11) reflection was used. The energy of the incident beam and the Bragg angle of the backscattering were 21.747 keV and approximately 89.98 deg, respectively. The scattered X-ray photons were collected by twelve spherical Si analyzers at the end of the 10 m horizontal arm, and counted by CdZnTe detectors. The energy resolution of the spectrometer $R(E)$ was determined from the IXS spectrum of a polymethyl methacrylate, and it was approximately 1.5 meV FWHM (depending on the analyzer crystals). The momentum transfer ($Q = 4\pi E \sin(\theta)/hc$, where $E$ and $\theta$ is photon energy and Brag angle, respectively) resolution was about 0.5 nm$^{-1}$.

We made a sample cell for the IXS measurements at low temperatures. The cell with single-crystalline sapphire windows is made of aluminum, which was attached to a He-compression-type refrigerator to control the temperature. A sample thickness was 2 mm for this study. The IXS spectra were measured at eight Q values from 1.5 to 10.7 $nm^{-1}$. The scans were made over ±25 meV energy range at $Q < 6$ $nm^{-1}$, and ±35 meV at $6 < Q < 11$ $nm^{-1}$. These scans required 1.5 and 2 hours per a scan, respectively. We carried out 2 scans at each $\theta$ and each temperature to obtain spectra with good statistics. An empty cell measurement was also carried out at 293K and the spectra were used for background subtraction. The temperatures and compositions of the sample for IXS measurements are shown by red triangles in Fig. 2.

The obtained IXS spectra $I(Q,E)$ were analyzed using the standard Damped Harmonic Oscillator (DHO) model. The specific formula is as follows:

$$\frac{I(Q,E)}{\int I(Q,E)dE} = \int \frac{S(Q,E')}{S(Q)} R(E-E')\, dE' \tag{2}$$

$$\frac{S(Q,E)}{S(Q)} = \frac{1}{\pi}\left[I_0 \frac{z_0}{E^2 + z_0^2} + I_1 \frac{2z_1 {E_1}^2}{(E^2 - E_1^2)^2 + 4z_1^2\omega^2}\right].$$

$S(Q,E)$ is the dynamic structural factor of the sample, and $R(E)$ is the instrument resolution function, which was measured using a PMMA sample. $I_0$ and $z_0$ represent the intensity and width of quasi-elastic scattering, respectively, while $E_1$, $I_1$ and $z_1$ represent the energy, intensity, and width of inelastic scattering, respectively. The measured spectra were fitted to this function, and the parameters were optimized. In particular, the IXS sound velocity was determined from the slope of the $Q$-dependence of the longitudinal acoustic mode $E_1$.

III. RESULTS

## A. Ultrasonic Measurements

Figure 3 shows (a) the temperature dependence and (b) the methanol mole composition ($x_m$) dependence of the sound velocity $v_{US}$ obtained from ultrasonic measurements. The accuracy of the sound velocity is within approximately 1–3 [m/s], which is smaller than the size of the mark. For pure water ($x_m$ = 0.00) in Fig.3 (a), the $v_{US}$ is consistent with literature values [16] (green dashed line) across the entire temperature range, including 298 K where the calibration was performed. In addition, the temperature dependence of methanol ($x_m$ = 1.00) (Fig. 3(a)) and the composition dependence of the water–methanol mixture at 283 K and 293 K can be compared with the values reported in the literatures [17] and [5], respectively. Although some discrepancies are observed, they do not exceed approximately 1%, and we therefore conclude that the results obtained are sufficiently accurate for the purposes of this paper.

Examining the temperature dependence for each composition in Fig3(a), when the composition of methanol is 0.5 or higher, the $v_{US}$ decreases almost linearly with increasing temperature, exhibiting the typical temperature dependence of a liquid. On the other hand, for compositions below this threshold, this linearity is gradually lost, and the behavior where the $v_{US}$ decreases at low temperatures becomes apparent. In particular, water exhibits anomalous behavior where the $v_{US}$ increases despite rising temperature.

These anomalies in the $v_{US}$ observed in the low-methanol composition and low-temperature regions become clearer when examining the composition dependence in Fig.3(b). In the right half of the figure—the high-methanol composition region—the $v_{US}$ curve at each temperature varies almost linearly with composition, and these $v_{US}$ curves are arranged at nearly equal intervals. This demonstrates a very simple composition-temperature dependence. In contrast, examining the data in the left half—the low-methanol composition region—reveals more complex variations; peaks appear in composition dependence, and their positions change with temperature. However, it is possible to grasp the overall characteristics through a simple interpretation: the $v_{US}$ decreases as one moves toward the low-methanol composition (pure water) region at low temperatures. These characteristics of temperature and composition dependence are nearly identical to previous results on the $v_{US}$ of water–ethanol mixtures [6,7], and it is considered reasonable to adopt the simple interpretation—common to both water–methanol and water–ethanol mixtures—that the $v_{US}$ decreases as the temperature approaches the low-temperature region of pure water [7]. However, the papers [5,6] focuses on the peak observed at low alcohol compositions in the composition-dependent behavior and posits the existence of special structures of water and alcohol. We intend to refute the existence of such a structure using IXS sound velocity data presented in the following sections.

## B. Inelastic X-ray Scattering Measurements

Figure 4 shows the spectral data obtained from IXS measurements at $x_m$ = 0.2, T = 273 K, and Q

= 2.91 nm$^{-1}$. The black circles represent the experimental data, the red line shows the fit using the model function as Eq. (1). The blue and green lines correspond to the first term (quasi-elastic scattering) and second term (inelastic scattering = longitudinal acoustic mode) of the model function in Eq. (1) without the resolution broadening, respectively. The pink line in the figure above represents the residuals between the experimental values and the fits, normalized by the measurement errors. Since this residual does not becomes large at any specific energy position and its absolute value is generally less than 1, we can conclude that the model function used here generally reproduces the experimental data well.

Figure 5(a) shows the $Q$-dependence of the energy of the obtained acoustic mode $E_1$ and Fig. 5(b) shows the dynamic sound velocity calculated by $v(Q) \equiv E_1/Q$. The dispersion relation for the longitudinal sound velocity is generally nearly linear at low $Q$ values, and with increasing $Q$, it typically begins to decrease around $Q_1/2$, where $Q_1$ is the first peak of $S(Q)$. In water-methanol mixtures, $Q_1$ is approximately 20 nm$^{-1}$ [18], and the dispersion relation observed here is consistent with this. As shown in Fig. 5(b), $v(Q)$ remains nearly constant, although it exhibits slightly lower values in the low $Q$ region below 2 nm$^{-1}$ and the high $Q$ region above 9 nm$^{-1}$. The decrease in the high-$Q$ region is due to the reason mentioned above. The decrease in the low-$Q$ region is thought to be because $v(Q)$ is asymptotically approaching $v_{US}$ —the value at the $Q$=0 limit— though it may also be due to the accuracy of the model function [19]. In any case, the IXS sound velocity $v_{IXS}$ was calculated as the average value over the range 2.0 < $Q$ < 8.5 nm, where $v(Q)$ remained nearly constant, and the variation in $v(Q)$ was used as the error bar.

The temperature and composition dependencies of $v_{IXS}$ in water-methanol mixtures obtained in this manner are shown as open circles with errorbars in Figs 6(a) and (c), respectively. Additionally, $v_{US}$ is plotted as lines on the same figures ($v_{US}$ values for $x_m$=0.5 and 0.75, which were not measured directly, were calculated by linear interpolation based on the composition. Since this composition range exhibits good linearity in temperature-composition dependence, it can be said that the error resulting from interpolation is not significant). What is immediately apparent is that $v_{IXS}$ takes on values significantly larger than $v_{US}$, indicating a so-called "fast sound" state[20,21]. In other words, it can be said that this system universally exhibits a strong frequency dependence of the sound velocity, that is, there exist dynamic fluctuations. This holds true not only for mixtures but also for water alone and for methanol alone; it becomes clear that there are other forms of "fluctuations" besides those caused by mixing of water and alcohol. Looking at the temperature dependence in Fig. 6(a), $v_{IXS}$ exhibits an almost linear change with temperature at all compositions. This result is significant in contrast to the fact that $v_{US}$ exhibits complex temperature dependence at low alcohol compositions. The bare sound velocity $v_{IXS}$ with mesoscopic fluctuations frozen out, can be said to exhibit "normal" temperature dependence. In Fig.6(b), the composition dependence of $v_{IXS}$ also shows a nearly linear variation at all temperatures: The bare sound velocity in a mixture is represented by the average of the sound velocities of water and methanol is something "normal" behavior. The nature of the bare sound velocity will be discussed further in the discussion section.

### C. Strength of dynamic fluctuation

As seen in the previous section, even at the same sound velocity, low-frequency ($v_{US}$) and high-frequency ($v_{IXS}$) waves exhibit completely different behavior. To quantify this difference, we plot the temperature and composition dependencies of strength of dynamic fluctuation $\Delta_v$ calculated by equation (1), in the lower panels (b) and (d) in Fig. 6. Examining the temperature dependence in (b), it can be seen that for $x_m$ = 0.0–0.2, $\Delta_v$ increases as the temperature decreases. In contrast, for $x_m$ > 0.5, the temperature dependence is relatively small. Furthermore, examining the composition dependence in (d) reveals that, at all temperatures, smaller values of $x_m$ correspond to larger values of $\Delta_v$, with the increase being particularly pronounced for $x_m$ ≤ 0.2. These results imply that the primary origin of dynamic fluctuation in the low methanol composition region lies not in the water–methanol mixture itself, but in water alone at low temperatures. In any case, it has become clear that this fluctuation is the origin of the maximum observed in the composition dependence of $v_{US}$ in water-methanol mixtures. It also should be noted that the value for methanol alone $x_m$=1.0 is not zero, which indicates that mesoscopic fluctuation exists even in pure methanol.

### D. Other water-alcohol mixtures

In addition to the water–methanol mixtures, we performed IXS measurements on water–ethanol and water–glycerol [14] mixtures at around room temperature to determine the composition dependence of $\Delta_v$. The results are shown in Figs. 7 and 8, respectively. When examining the composition ($x_e$) dependence of $v_{IXS}$, it shows an almost linear variation in the water-ethanol mixtures (Fig. 7(a)). Furthermore, in the water–glycerol mixtures shown in Fig. 8(a), $v_{IXS}$ remains nearly constant regardless of composition ($x_g$), which can also be recognized as an almost linear dependence. Examining the composition dependence of $\Delta_v$, it is observed that in both the water–ethanol mixtures (Fig. 7(b)) and the water–glycerol mixtures (Fig. 8(b)), $\Delta_v$ is not zero at any composition, and in each case, an increase toward pure water is observed in the low-composition region ($x_e \lesssim 0.15$, $x_g \lesssim 0.2$). These results are similar to those for water–methanol mixtures and are independent of whether the composition dependence of $v_{US}$ exhibits a maximum (methanol and ethanol) or not (glycerol). In the end, the following conclusion can be drawn for all three mixture systems: Significant fluctuation originating from the supercooled region already exist in pure water; Although this fluctuation is dominant in the mixture systems as well, it gradually decreases when alcohol is added or the temperature is raised, and generally disappear at around 15–20% alcohol composition.

## IV. DISCUSSION

Here, we will further explore the thermodynamic anomalies in water–alcohol mixtures by examining the experimental results obtained, while considering the physical significance of each parameter appearing in Eq. (1). (i) $v(\infty) \approx v_{IXS}$: The bare sound velocity at the microscopic level, unaffected by mesoscopic fluctuations. According to the one-dimensional oscillator model that describes sound wave (phonon), the sound velocity is expressed solely in terms of the strength of interparticle interactions (force constant), the interparticle distance, and the mass of the particle [23]; thus, it can be regarded as a physical quantity that reflects the "microscopic stiffness" of a substance, which is determined solely by its local structure. It is fundamentally not a thermodynamic quantity and does not correlate directly with other thermodynamic quantities. In fact, the experimental results for the water-alcohol mixtures show that the temperature dependence of $v_{IXS}$ exhibits an almost linear change with temperature for all compositions, indicating the "normal" result that the microscopic stiffness becomes monotonically decreases as the temperature rises. This is unrelated to the anomalous temperature dependence of various thermodynamic quantities such as $v(0)$. Furthermore, the composition dependence also exhibits a nearly linear variation, suggesting that in the mixture, the microscopic stiffness of water molecules and that of methanol molecules are simply averaged. While we cannot rule out the existence of any unique rigid local structures, as has been discussed in the past regarding this system, or behavior involving complex interactions between hydrogen bonds between water and alcohol, the results do not support such hypotheses. This, too, can be considered a "normal" result, unrelated to the anomalous behavior observed in the composition dependence of thermodynamic quantities, such as excess specific heat and excess enthalpy, where maxima and minima are observed [3].

(ii) $\Delta_v$: This represents the mesoscopic-level fluctuation strength. In this study, we found that, across all water–alcohol (methanol, ethanol, and glycerol) mixtures, pure water exhibited the highest fluctuation strength, which then decreased with increasing alcohol composition in the low-composition range. This leads to the seemingly counterintuitive conclusion that mixing substances reduces fluctuations. On the other hand, this result is consistent with the liquid–liquid phase transition (LLT) scenario of water [24,25,1], which comprehensively explains the thermodynamic anomalies of water — namely, that liquid water actually consists of two phases, a high-density and a low-density phases, and that in the supercooled region, LLT occurs, similar to a discontinuous phase transition such as LGT, and a LLT critical point exists as the endpoint—the existence of such LLT critical fluctuation has been observed in our measurements of dynamic fluctuations in water over a wide range of temperature and pressure conditions above room temperature [1], and it is expected that this fluctuation will become even larger at low temperatures. Although the existence of this LLT critical point has not yet been proven, the present results can, on the contrary, be regarded as strong evidence for its existence. Furthermore, simulation results have shown that such LLT exists not only in water but also in the much lower temperature region of methanol [26]. In other words, to understand the state of fluctuations in water–alcohol mixtures, one must consider not only the effects of mixing the two substances but also the fluctuations inherent in water alone and alcohol alone. The fact that fluctuations were observed across a wide range of composition and temperature ranges in this study is thought to be due to this reason. Ideally, it would be desirable to separate fluctuations of different origins, but this is currently difficult.

(iii) $v(0) \approx v_{US}$: While defined as the velocity at which sound waves propagate on a macroscopic scale, it also functions as a thermodynamic quantity, as evidenced by the relationship between macroscopic density $\rho$ and macroscopic adiabatic compressibility $\beta_S$ as $v(0) = (\rho\beta_S)^{-0.5}$. It is a physical quantity representing the "macroscopic stiffness" of a substance, and for many substances, it coincides with the "microscopic stiffness", $v(0) \approx v(\infty)$. This relationship is also indicated in the solid-state textbook [23]. However, the present results indicate that, at least for liquid water, alcohol, and mixtures of these, this equality does not hold. The macroscopic thermodynamic quantity $v(0)$ is determined not only by the microscopic $v(\infty)$ but also the mesoscopic $\Delta_v$. In these results, while $v(\infty)$ exhibits a "normal" temperature-composition relationship, the presence of "anomalies" in $\Delta_v$ allows us to conclude that an anomaly has appeared in the thermodynamic quantity $v(0)$. In this way, it has often been common practice to separate and discuss thermodynamic anomalies in liquids into "normal" and "anomalous" terms; in the past ultrasonic measurements of water–ethanol mixtures, the temperature-composition dependence of the macroscopic compressibility $\beta$ was also discussed by separating it into normal $\beta^n$ and anomalous $\beta^{an}$ components as, $\beta = \beta^n + \beta^{an}$; Here, it is also assumed that the anomalous term originates from water alone [7]. While the present discussion is similar to this one, unlike the aforementioned paper —where the separation of these two terms was based on an arbitrary assumption— the separation of the normal and anomalous terms in this study is based on experimental results showing frequency dependence, and thus involves no arbitrariness. Furthermore, the physical interpretations of the two terms are well-defined, and it can be said that the thermodynamic anomalies of this system have become clearer.

## V. CONCLUDING REMARKS

We performed US and IXS measurements on a water-alcohol (methanol, ethanol and glycerol) mixtures, and determined the composition and temperature dependencies of the strength of dynamic fluctuation. The results revealed that dynamic fluctuations are universally present in all systems and across all composition and temperature ranges. Furthermore, in all systems, an increase in the strength of dynamic fluctuation was observed as the alcohol composition decreased within the low-composition range of 15–20 %, with pure water exhibiting the greatest strength. In the water–methanol mixture, temperature dependence was also measured, and this increase was greater at lower temperatures. In other words, common to all three systems, this indicates that in such low-composition regions, fluctuation originating from the low-temperature (supercooled) region of water alone is far larger and more dominant than those caused by the mixing of water and alcohol. It is known that various anomalies appear in the temperature dependence of the $v_{US}$ in water–methanol and water–ethanol mixtures; it has now been proven that these anomalies originate from the fluctuation in water alone. This fluctuation in water is consistent with the LLT scenario for water, which posits the existence of an LLT critical point in the supercooled region, and are also consistent with the results of our previous measurements of dynamic fluctuations in water[1]. This can be considered one of the strongest pieces of evidence for the existence of the LLT critical point, which has not yet been proven.

It has long been known that water-alcohol mixtures exhibit unique thermodynamic properties, and the origin of these properties has been debated in terms of specific mixing states at certain compositions or the presence of unique local structures. It has also been suggested that the origin of these properties varies depending on the type of alcohol. However, these conventional arguments were based on the implicit assumption that the individual components—water or alcohol—were homogeneous prior to mixing. However, the results of the dynamic fluctuation measurements in this study indicate that the assumption of homogeneity in the individual substances prior to mixing is incorrect. Furthermore, since almost no anomalies were observed in the microscopic sound velocity obtained by IXS, the results cast doubt on the existence of unique local structures. On the other hand, by focusing on mesoscopic-level fluctuations, we were able to arrive at a comprehensive interpretation that is independent of the specific alcohol species. We believe we have identified a new direction for understanding the thermodynamics of liquids. We look forward to further research focusing on the mesoscopic level, including theoretical studies, simulations, and small-angle scattering—a conventional method for measuring density fluctuations.

## ACKNOWLEDGEMETS

This work was partially supported by JSPS (Japan Society for the Promotion of Science) KAKENHI Grant Numbers JP24740265 and JP26610116. The synchrotron radiation experiments were performed at BL35XU of SPring-8 with the approval of the Japan Synchrotron Radiation Research Institute (JASRI) (Proposal Nos. 2013A1397, 2014B1365, and 2015A1466). The authors are grateful to Emeritus Professor Yoshimi Tsuchiya for providing us with the ultrasonic measurement system and for his kind advice regarding its use. The authors would also like to thank S. Tsutsui and A. Q. R. Baron for their cooperation with the IXS experiments.

## AUTHOR DECLARATIONS

### Conflict of Interest

The authors have no conflicts to disclose.

### Author Contributions

**Yukio Kajihara**: Conceptualization and Supervision; Investigation(equal); Formal Analysis(equal); Writing-review and editing(lead). **Nanako Shibata:** Investigation(equal); Formal Analysis(equal); Writing-original draft. **Masanori Inui:** Investigation (equal); Writing-review and editing(supporting)

## DATA AVAILABILITY

The data that support the findings of this study are available from the corresponding author upon reasonable request.

FIG.1

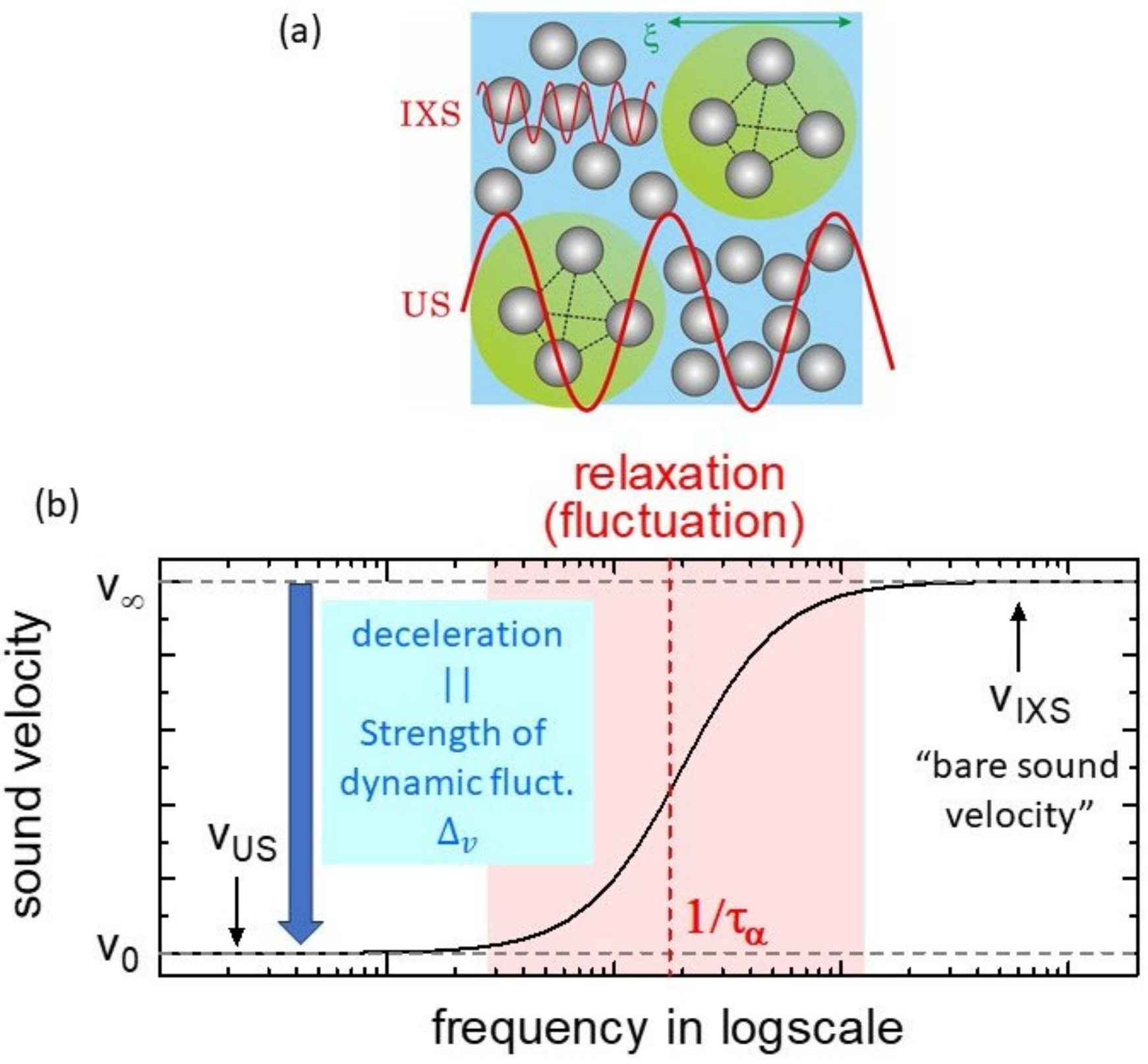

Schematic diagram of the principle of dynamic fluctuation measurement. (a) If mesoscopic-level fluctuations or heterogeneities exist in a sample, the sound velocity at sufficiently low frequencies (long wavelengths) such as US is reduced, whereas the sound velocity at sufficiently high frequencies (short wavelengths) as measured by IXS is not reduced. The latter can be recognized as the "bare sound velocity" unaffected by fluctuations, and is thought to directly reflect the stiffness of the local structure. (b) The effect of fluctuations manifests as a frequency dependence of the sound velocity. Two sound-velocity-measurement methods with vastly different frequencies such as IXS and US, the degree of deceleration is extracted as the strength of dynamic fluctuation $\Delta_v$.

FIG.2

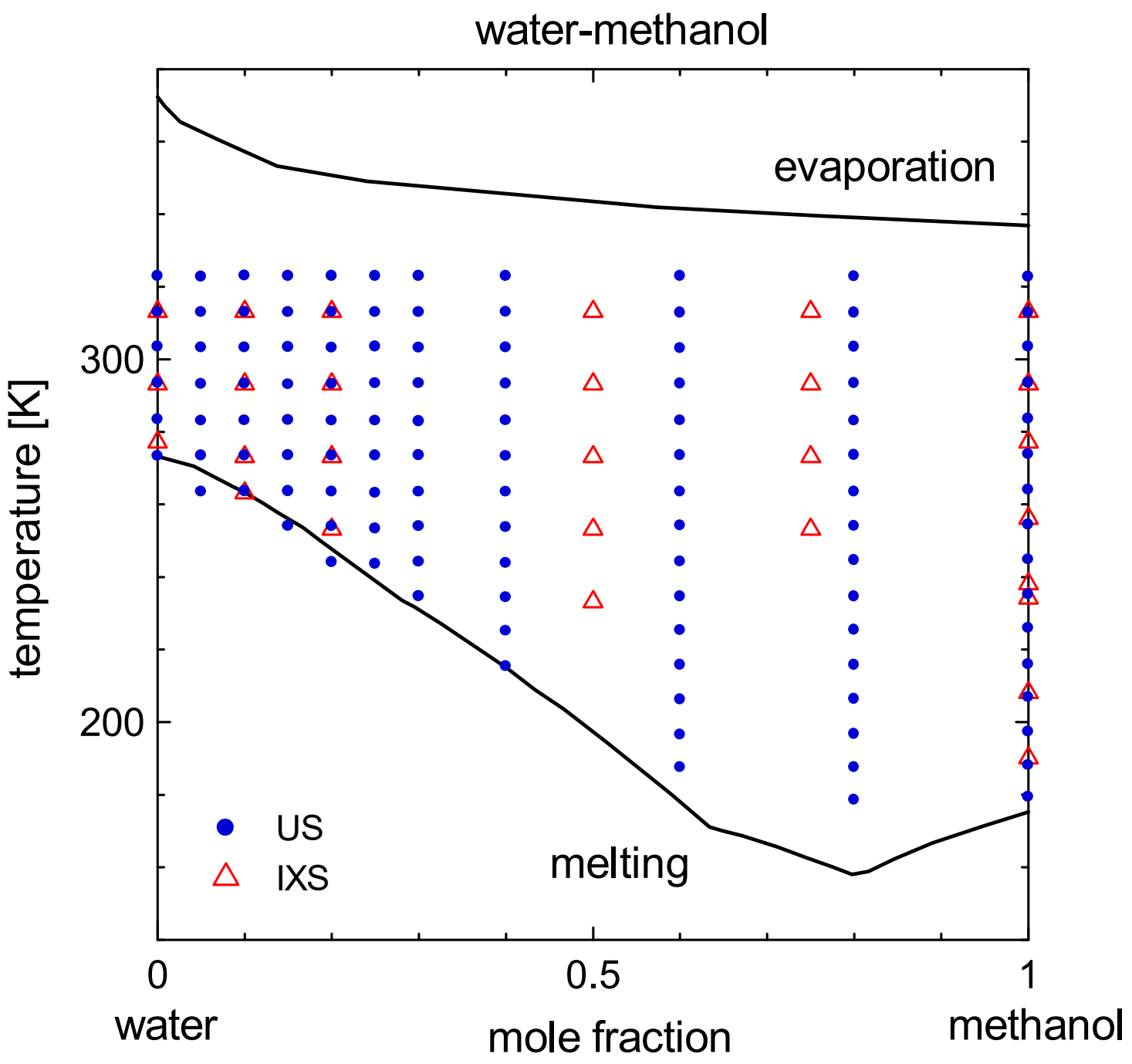

Temperature-composition diagram for sound velocity measurements in water-methanol mixtures.

Blue circles indicate US experiments; red triangles indicate IXS experiments.

FIG.3

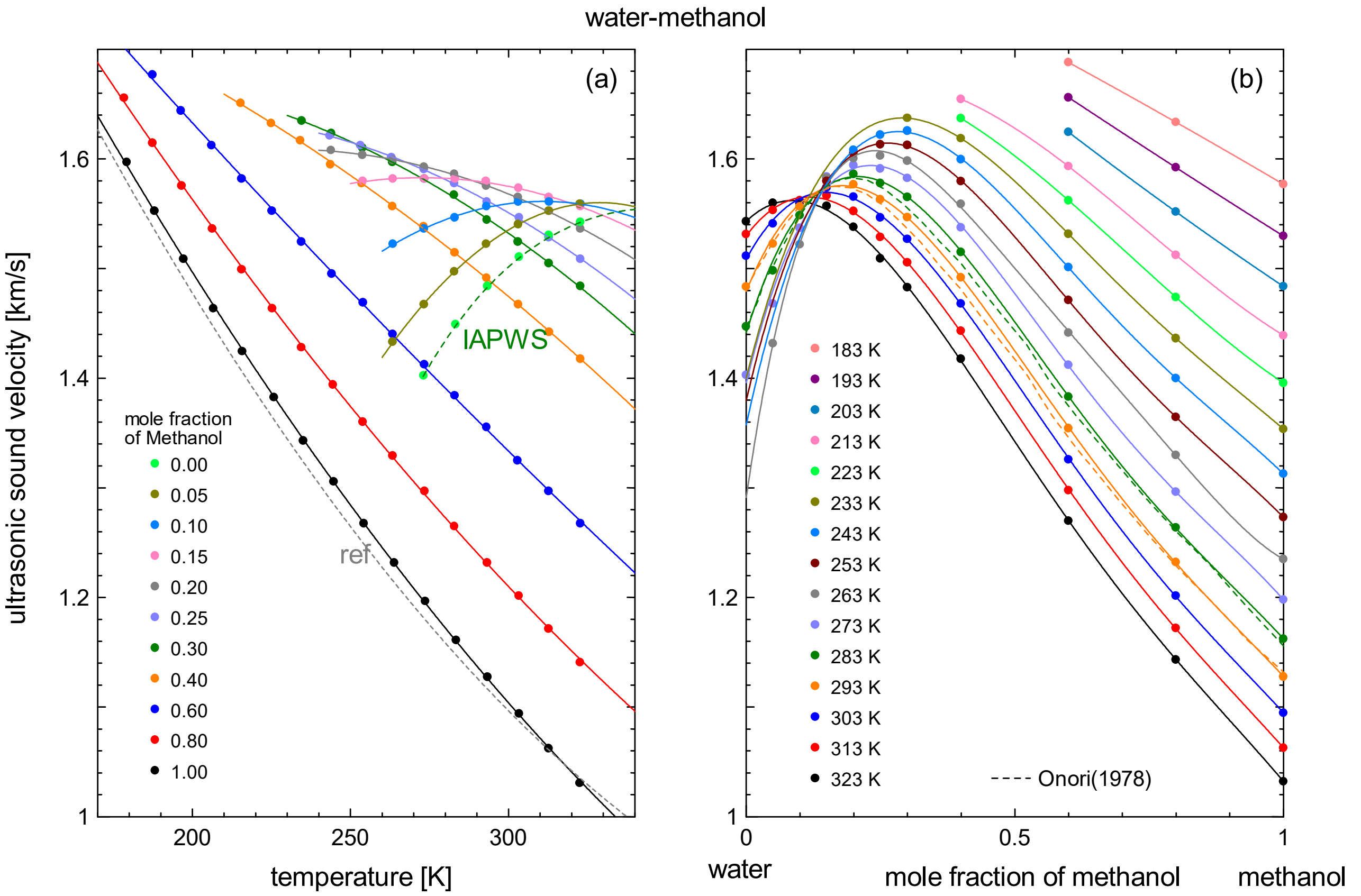

(a) Temperature dependence and (b) composition dependence of the ultrasonic sound velocity in water–methanol mixtures. The dashed lines represent literature values for water [16], methanol [17] and water-methanol mixtures [5].

FIG.4

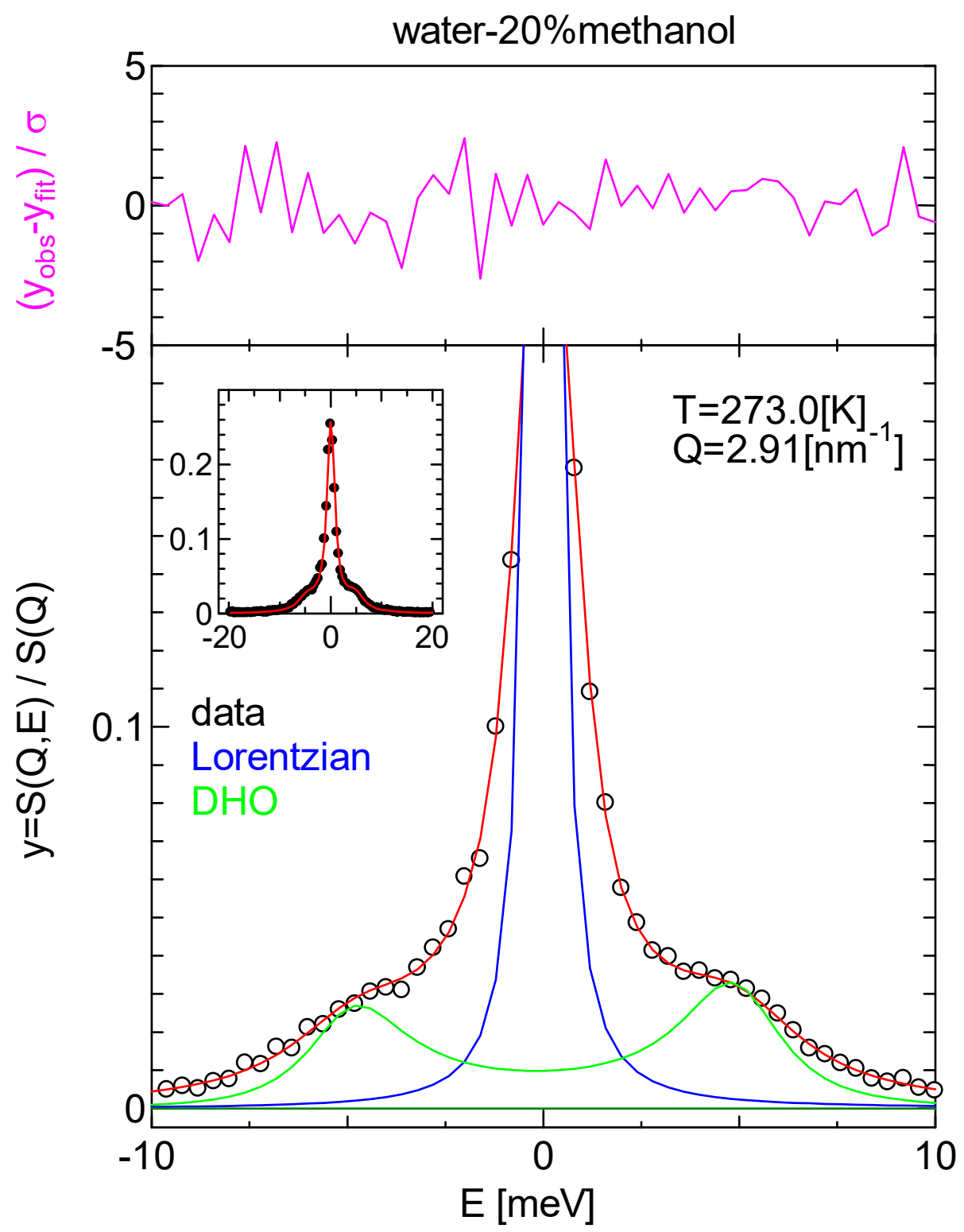

(below) A typical IXS spectrum with a methanol composition of 20%, a temperature of 273 K, and a Q value of 2.91 $nm^{-1}$. The circles represent experimental data, and the red line shows the fit obtained using the model function. The blue and green lines correspond to the first and second terms of Eq. (1), respectively; broadening due to the instrument's resolution has not been taken into account. (above) Residuals between the experimental data and the fit.

FIG.5

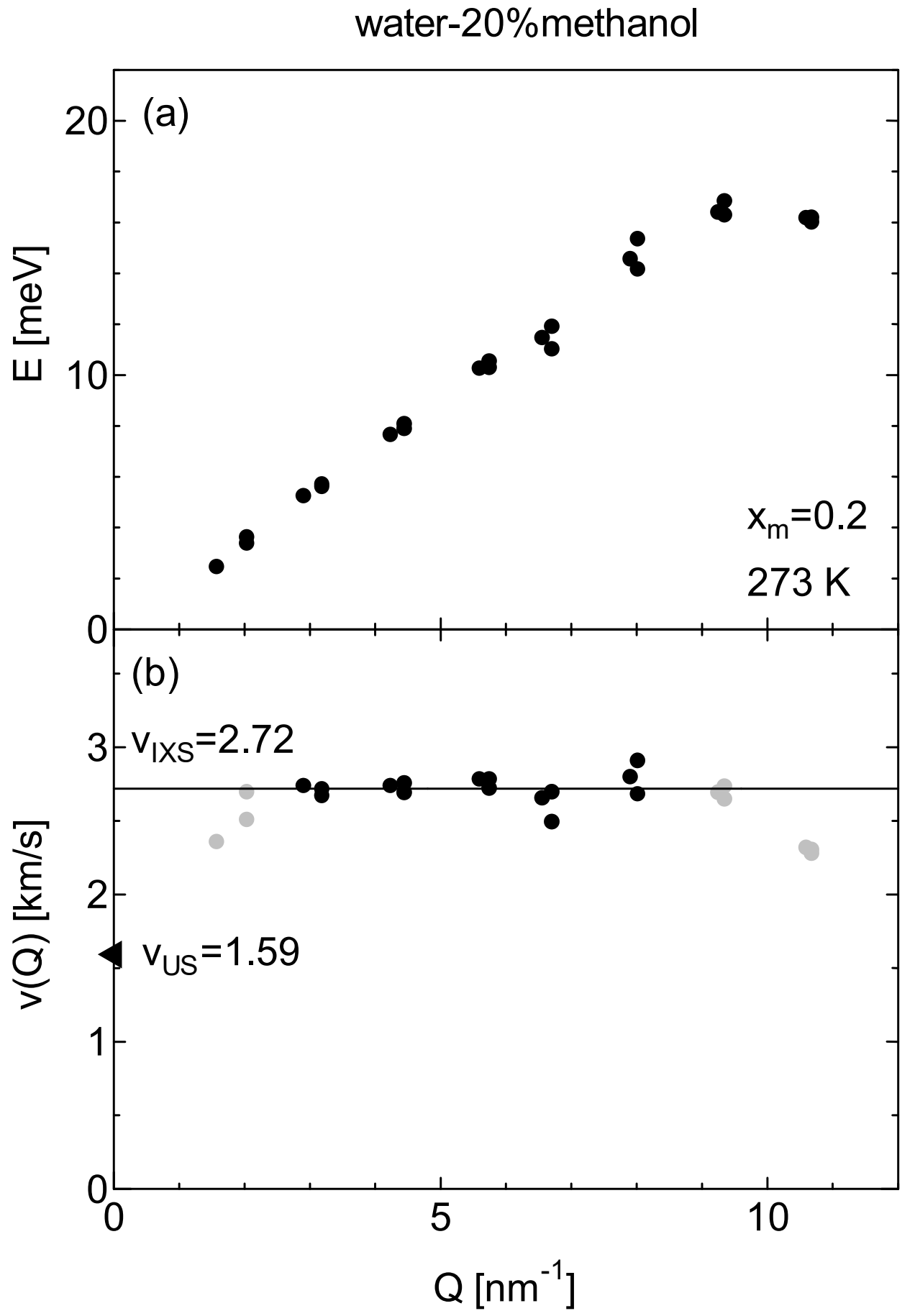

A typical (a) dispersion relation ($Q$-dependences of estimated energy of inelastic mode) and (b) dynamic sound velocities $v(Q)$ for a methanol composition of 20% and a temperature of 273 K. $v_{IXS}$ was determined using the average value of $v(Q)$ over the $Q$ range ($2.5 < Q < 8.5$ nm$^{-1}$) where the $Q$-dependence remained nearly constant. $v_{US}$ is also indicated by triangle at $Q$=0.

FIG6.

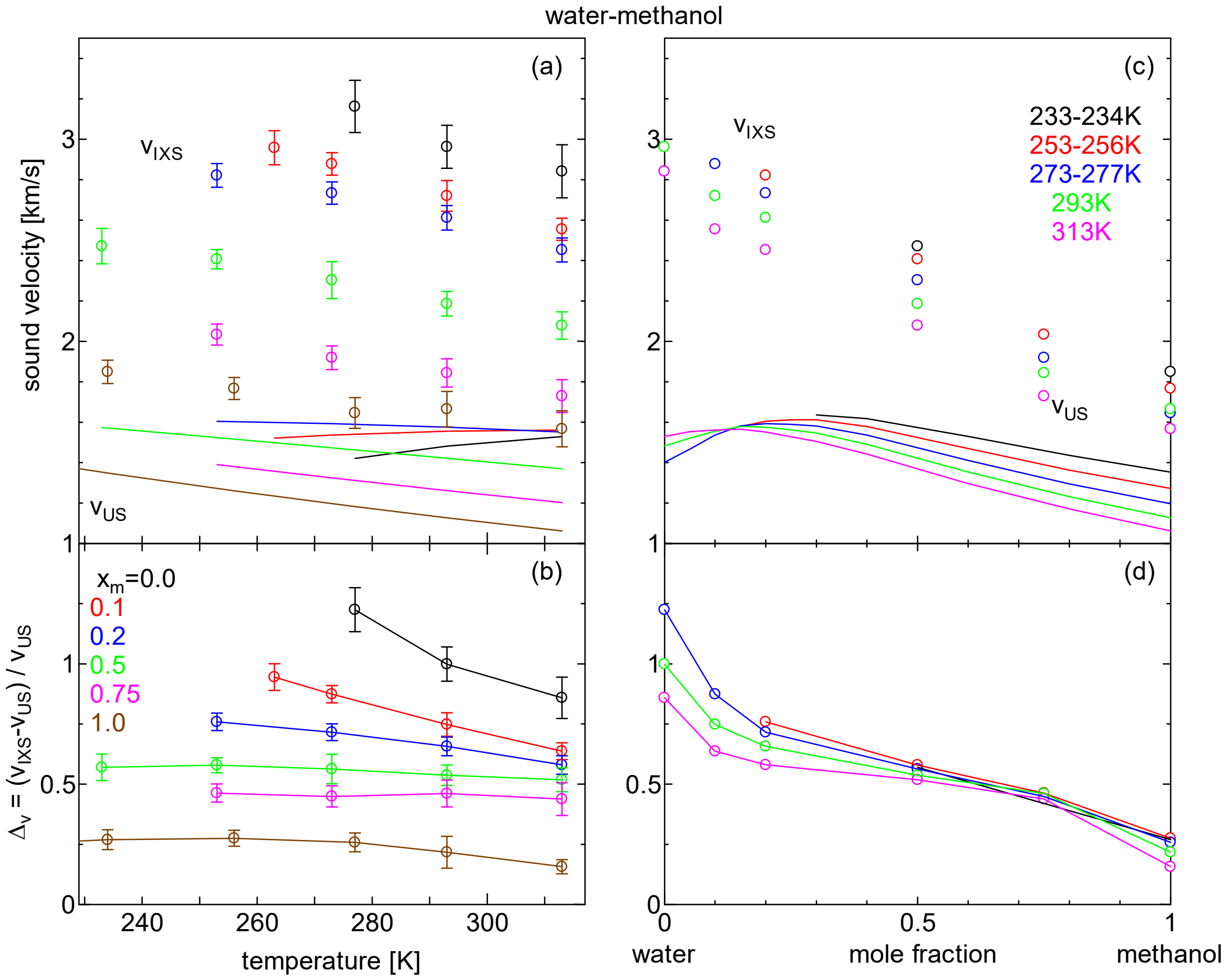

(a) Temperature dependence and (c) composition dependence of $v_{IXS}$ in water-methanol mixtures (circles). The $v_{US}$ are also shown as solid lines. Additionally, the (b) temperature dependence and (d) composition dependence of $\Delta_v$ calculated from $v_{IXS}$ and $v_{US}$ are plotted as circles in the figure below. The solid lines are visual guides.

FIG.7

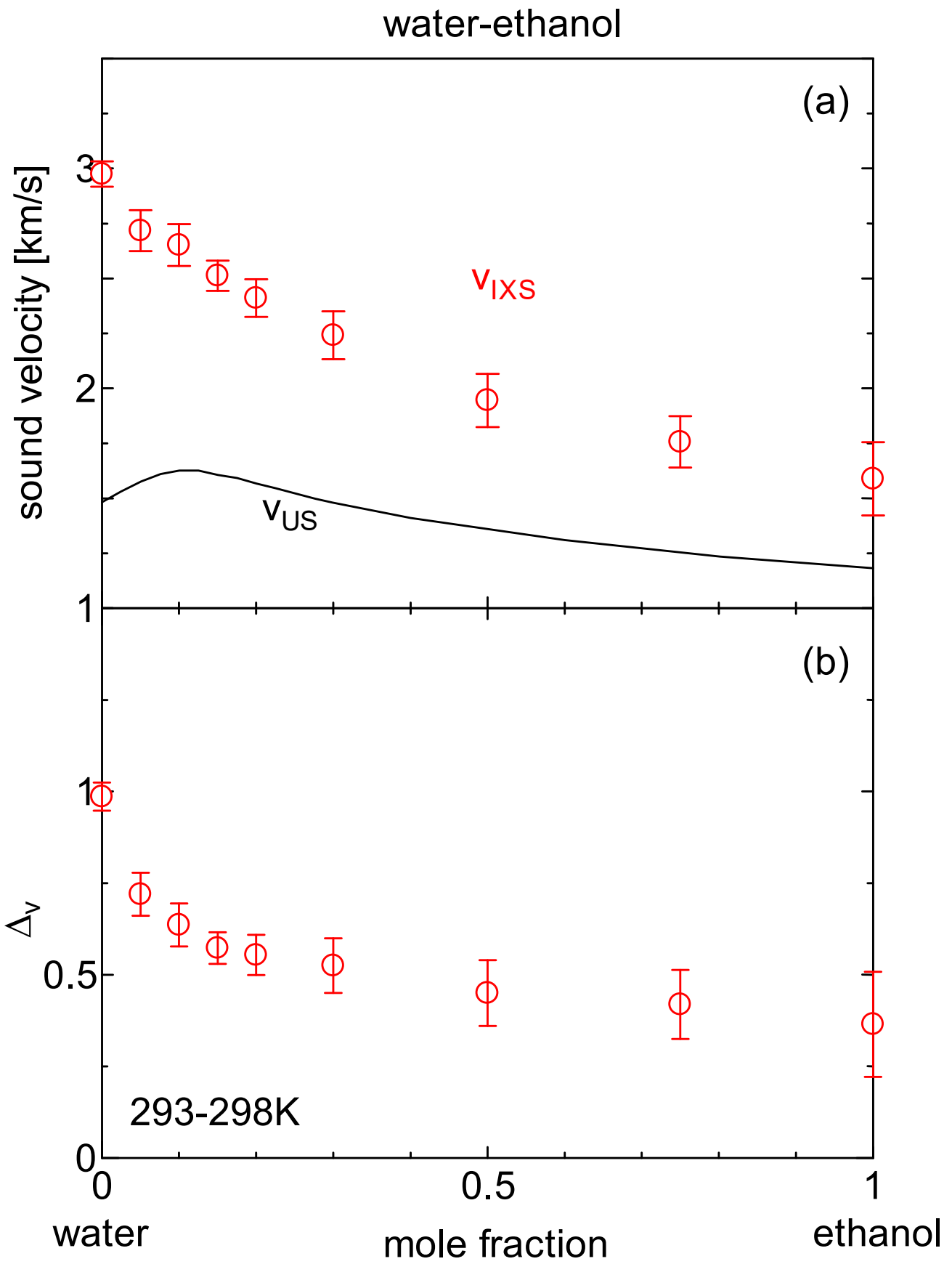

Composition dependence of (a) $v_{IXS}$ (red marks) and $v_{US}$ (black line) [7], and (b) $\Delta_v$ for water–ethanol mixtures.

FIG. 8

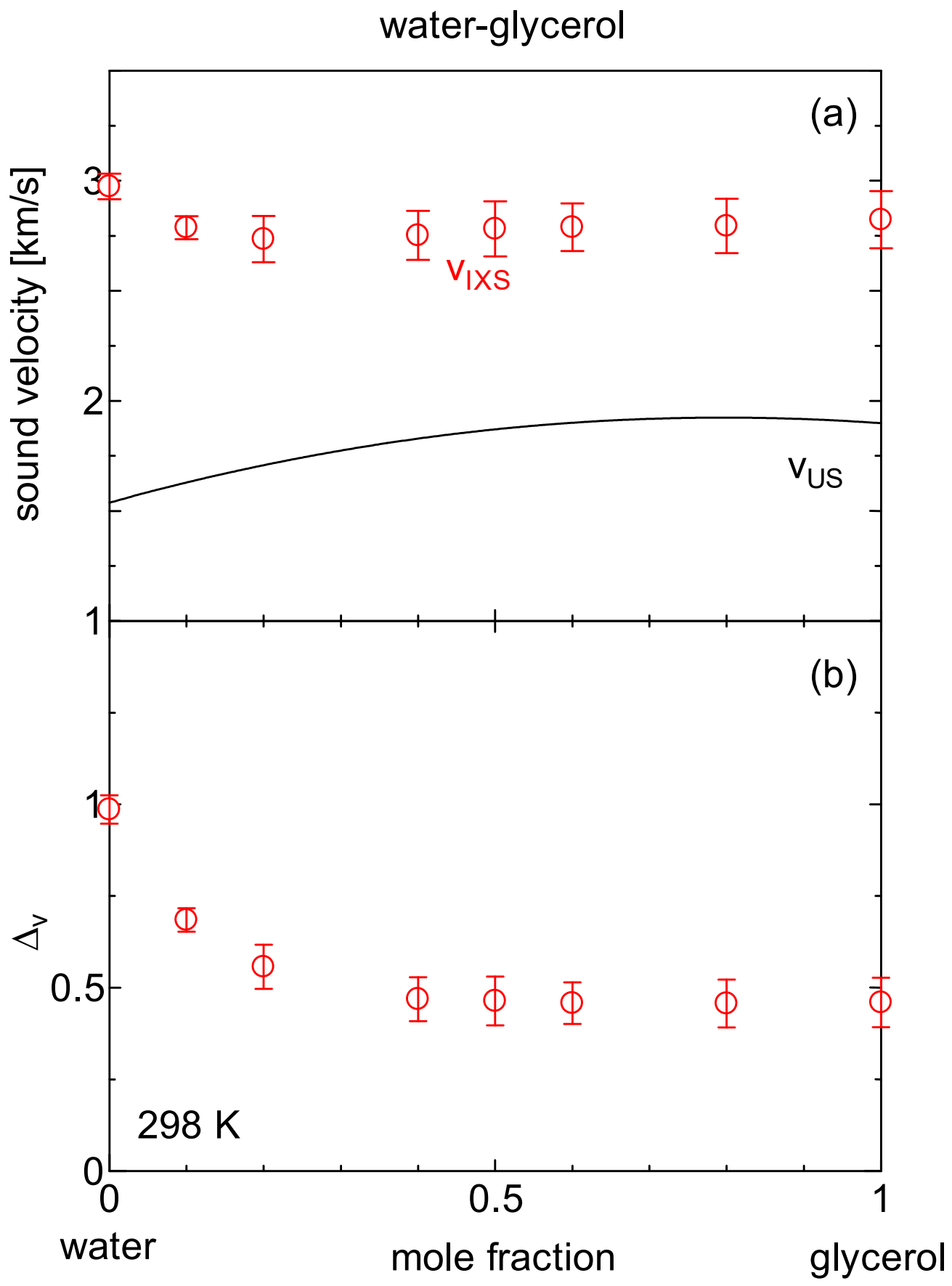

Composition dependence of (a) $v_{IXS}$ (red marks) and $v_{US}$ (black line) [22], and (b) $\Delta_v$ for water–glycerol mixtures.